\documentclass{elsart}
\usepackage{amsmath,bm}
\usepackage{graphicx}

\newcommand{\Msun}{$\mathrm{M}_{\odot}$}

\usepackage[normalem]{ulem}  
\usepackage[dvips]{color} 

\begin{document}

\begin{frontmatter}

\title{Constraining the Radii of Neutron Stars with Terrestrial
Nuclear Laboratory Data}

\author{Bao-An Li}
\ead{Bao-An\_Li@Tamu-Commerce.edu}
\address{Department of Physics, Texas A\&M University-Commerce,
P.O. Box 3011, Commerce, TX 75429-3011, USA\\
and Department of Chemistry and Physics, P.O. Box 419, 
Arkansas State University, State
University, Arkansas 72467-0419, USA} 
\author{Andrew W. Steiner}
\address{Theoretical Division, Los Alamos National Laboratory, Los
Alamos, NM 87545, USA}

\begin{abstract}
Neutron star radii are primarily determined by the pressure of isospin
asymmetric matter which is proportional to the slope of the nuclear
symmetry energy. Available terrestrial laboratory data on the isospin
diffusion in heavy-ion reactions at intermediate energies constrain
the slope of the symmetry energy. Using this constraint, we show that
the radius (radiation radius) of a 1.4 solar mass (\Msun) neutron star
is between 11.5 (14.4) and 13.6 (16.3) km.
\end{abstract}

\begin{keyword}
\PACS 25.70.-z \sep 26.60.+c \sep 97.60.Jd \sep 24.10.-i
\end{keyword}

\end{frontmatter}

With central pressures of $10^{36}$ dynes/cm$^2$ and gravitational
binding energies of $10^{53}$ ergs, neutron stars are among the most
exotic objects in the universe. Impressive progress has been made on
the observable properties of neutron stars, such as masses, radii,
spectra, and rotational
properties~\cite{thor99,haen01,bob02,bob02b,gendre02a,gendre02b,ryb05,becker}.
However, a precise neutron star radius measurement still eludes us.

The theoretical understanding of these observable properties demands
an understanding of the relevant nuclear physics. For recent reviews,
see
Refs.~\cite{lat01,lat01b,lat01c,prak01,pethick04,haen03,henning00,henning00b,steiner05a}.
The global properties of neutron stars: masses, radii, and
composition, are determined by the Equation of State (EOS) of
neutron-rich nucleonic matter, thus neutron stars are ideal
astrophysical laboratories for investigating the EOS. The EOS can be
separated into two contributions, the isospin symmetric part (the EOS
of nuclear matter, $E_{\mathrm{nuc}}$) and the isospin asymmetric part
(the nuclear symmetry energy, $E_{\mathrm{sym}}$). This separation is
manifest in the relation
$E(\rho,\delta)=E_{\mathrm{nuc}}(\rho)+\delta^2
E_{\mathrm{sym}}(\rho)$, where $\rho$ is the baryon density,
$\delta=(\rho_n-\rho_p)/\rho$ is the isospin asymmetry, and $\rho_n$
and $\rho_p$ are the neutron and proton densities. While many neutron
star properties depend on both parts of the equation of state, the
radius is primarily determined by the slope of the symmetry energy,
$E_{\mathrm{sym}}^{\prime}(\rho)$~\cite{lat01,lat01b,lat01c,prak01}.
Unfortunately, our knowledge about the density dependence of the
nuclear symmetry energy has been rather poor. Predictions of the
symmetry energy by nuclear many-body theories vary
significantly~\cite{diep}. Because of its importance for neutron star
structure, determining the density dependence of the symmetry energy
has been a major goal of the intermediate energy heavy-ion
community. Although extracting the symmetry energy is difficult
because of the complicated role of isospin in the reaction dynamics,
several observable probes of the symmetry energy have been
suggested~\cite{li97,li00x,li02x,lireview} (see also
Refs.~\cite{libookb,pawel,baran05} for reviews).

Some significant progress has been made recently in determining the
density dependence of $E_{\mathrm{sym}}(\rho)$ using: (i) isospin
diffusion in heavy-ion reactions at intermediate energies as a probe
of the $E_{\mathrm{sym}}(\rho)$ around the saturation
density~\cite{shi,betty04,chen05a,steiner05b,lichen,chen05b}, (ii)
flow in heavy-ion collisions at higher energies to constrain the
equation of state of nuclear matter~\cite{pawel}, and (iii) the sizes
of neutron skins in heavy nuclei to constrain $E_{\mathrm{sym}}(\rho)$
at sub-saturation densities~\cite{chuck,chuck2,steiner05a,jorge}.

The observational determination of a neutron star radius from the
measured spectral fluxes relies on a numerical model of the neutron
star atmosphere and uses the composition of the atmosphere, a
measurement of the distance, the column density of x-ray absorbing
material, and the surface gravitational redshift as inputs. Many of
these quantities are difficult to measure, thus the paucity of radius
measurements. Current estimates obtained from recent x-ray
observations have given a wide range of results.

In this Letter, we combine recently obtained isospin diffusion data,
information from flow observables, studies on the neutron skin of
$^{208}$Pb, and other information to constrain the radius of $1.4$
\Msun~neutron stars.

We use the EOS corresponding to 
the potential~\cite{das03,ibuu04,ibuu04b}
\begin{eqnarray}
U(\rho ,\delta ,\vec{p},\tau ,x) &=&A_{u}(x)\frac{\rho _{\tau ^{\prime
}}}{\rho _{0}}+A_{l}(x)\frac{\rho _{\tau }}{\rho _{0}} \notag \\
&+&B(\frac{\rho }{\rho _{0}})^{\sigma }(1-x\delta ^{2})-8\tau
x\frac{B}{\sigma +1}\frac{\rho ^{\sigma -1}}{\rho _{0}^{\sigma
}}\delta \rho _{\tau ^{\prime }} \notag \\ &+&\frac{2C_{\tau ,\tau
}}{\rho _{0}}\int d^{3}p^{\prime }\frac{f_{\tau }(%
\vec{r},\vec{p}^{~\prime})}{1+(\vec{p}-\vec{p}^{~\prime })^{2}/\Lambda
^{2}} \notag \\ &+&\frac{2C_{\tau ,\tau ^{\prime }}}{\rho _{0}}\int
d^{3}p^{\prime }\frac{f_{\tau ^{\prime }}(\vec{r},\vec{p}^{~\prime
})}{1+(\vec{p}-\vec{p}^{~\prime })^{2}/\Lambda ^{2}}.  \label{mdi}
\end{eqnarray}
Here $\tau =1/2$ ($-1/2$) for neutrons (protons) with $\tau \neq \tau
^{\prime }$, $\sigma=4/3$, and $f_{\tau }(\vec{r},\vec{p})$ is the
phase space distribution function. The incompressibility $K_{0}$ of
symmetric nuclear matter at $\rho _{0}$ is set to be $211$
MeV. Eq.~\ref{mdi} is an extension of the potential from Welke et
al.~\cite{mdyi,mdyib} from symmetric to asymmetric matter. The
parameter $x$ was introduced to mimic various predictions on
$E_{\mathrm{sym}}(\rho)$ by using different many-body theories and
effective interactions. It is a convenient way to parameterize
the uncertainty in the magnitude and density dependence of the
symmetry energy while keeping the isospin-symmetric part of the EOS
unchanged. The functions $A_{u}(x)$ and $A_{l}(x)$ depend on $x$
according to $A_{u}(x)=-95.98-x\frac{2B}{\sigma +1}$,and
$A_{l}(x)=-120.57+x\frac{2B}{\sigma+1}$ such that the same saturation
properties of symmetric matter and a value of
$E_{\mathrm{sym}}(\rho_0)=32$ MeV are obtained.

The isoscalar potential estimated from $(U_{neutron}+U_{proton})/2$
agrees very well with predictions from variational many-body
theory~\cite{wiringa}. In addition to being useful at the lower
energies discussed here, the underlying EOS has been tested
successfully against nuclear collective flow data in relativistic
heavy-ion reactions~\cite{mdyi,mdyib,pawel,zhang} for densities up to
five times saturation density. Also, the strength of the
momentum-dependent isovector potential at $\rho_0$ estimated from
$(U_{neutron}-U_{proton})/2\delta$ agrees very well with the Lane
potential extracted from nucleon-nucleus scatterings and (p,n) charge
exchange reactions with beam energies up to about 100
MeV~\cite{ibuu04,ibuu04b,data,data2}.

We focus on the properties of spherically-symmetric, non-rotating,
non-\-mag\-ne\-tized neutron stars at zero temperature by solving
the Tolman-Oppenheimer-Volkov equation. For the equation of
state below about 0.07 fm$^{-3}$, we use the results from
Refs.~\cite{nveos,bpseos}. Also, we assume that the neutron star
consists of $npe\mu$ matter, but does not contain any exotic
components, such as hyperons, quarks, or Bose condensates.

\begin{figure}[tbh]
\begin{center}
\includegraphics[scale=0.40,angle=0]{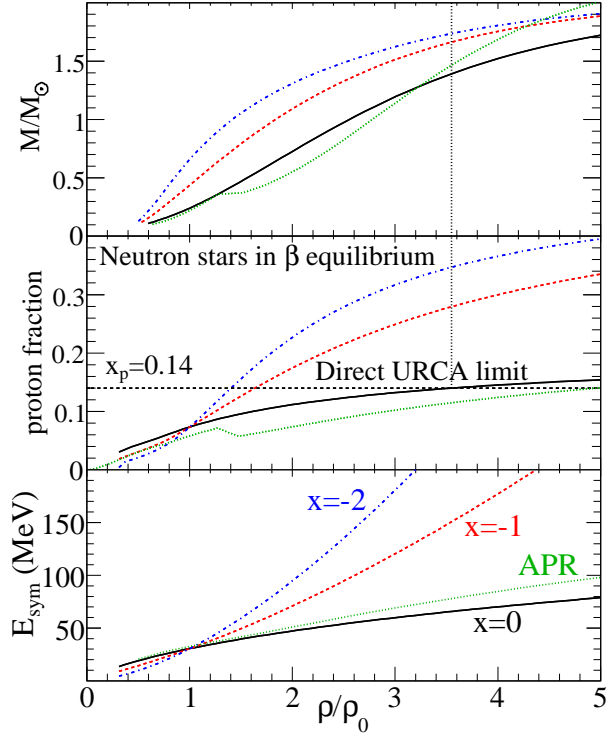}
\caption{{\protect\small Neutron star mass as a
function of the central density, and the proton fraction for
beta-equilibrated matter and symmetry energy as a function of density
for the EOS with $x=0,-1,$ and $-2$. The dotted lines give the
corresponding results for the APR EOS.}}
\end{center}
\label{figure1}
\end{figure}

In Fig.~1 we display some of the basic properties of the
EOS and the corresponding neutron stars.  Shown in the lower panel is
the symmetry energy for $x=0, -1$ and $-2$, respectively.  With $x=0$
the symmetry energy agrees very well with the prediction from
Akmal, et. al. (APR)~\cite{apr} up to about $5\rho_0$. Around $\rho_0$, the
$x=0$ EOS can be well approximated by $E_{\mathrm{sym}}^{x=0}(\rho)\approx
32(\rho/\rho_0)^{0.7}$. With $x=-1$, the $E_{\mathrm{sym}}^{x=-1}(\rho)\approx
32(\rho/\rho_0)^{1.1}$ is closer to predictions of typical
relativistic mean field models~\cite{steiner05a}. 

The middle panel shows the proton fraction, $x_p$, as a function of
density, and the top panel gives the mass of a neutron star as a
function of the central density. For $x_p$ below 0.14~\cite{lpph}, the
direct URCA process does not proceed because energy and momentum
conservation cannot be fulfilled. The proton fraction is sensitive to
the slope of the symmetry
energy~\cite{lat01,lat01b,lat01c,prak01}. For the $x=-1$ and $x=-2$
EOSs, the condition for direct URCA is fulfilled for nearly all
neutron stars above 1 \Msun. For the $x=0$ EOS, the minimum density
for direct URCA is indicated by the vertical dotted line, and the
corresponding minimum neutron star mass is indicated by the horizontal
dotted line. For the $x=0$ EOS, neutron stars with masses above 1.39
\Msun will have a central density above the threshold for the direct
URCA process.

Isospin diffusion in heavy-ion reactions is the re-distribution
process of isospin asymmetries carried originally by the colliding
partners. The degree and rate of this process depends on the relative
pressures of neutrons and protons, namely the slope of the
$E_{\mathrm{sym}}(\rho)$. It is harder for neutrons and protons to mix
up with a stiffer $E_{\mathrm{sym}}(\rho)$, leading to a
smaller/slower isospin diffusion.  Moreover, the distribution of the
isospin asymmetry versus density during heavy-ion reactions is
completely determined by the $E_{\mathrm{sym}}(\rho)$. As an
illustration, shown in the insert of Fig.\ 2 is a snapshot at 20 fm/c
of the correlation between the local isospin asymmetry and density in
the $^{124}$Sn+$^{124}$Sn reaction using the two extreme density
dependences of the $E_{\mathrm{sym}}(\rho)$. In this work, our 
calculations of nuclear reactions are performed using the
latest isospin and momentum-dependent transport model using in-medium
nucleon-nucleon cross sections consistent with the corresponding
single particle potential~\cite{lichen}. With the very stiff
symmetry energy of $x=-2$, a very neutron-rich dilute cloud surrounds
a more symmetric denser region up to $1.6\rho_0$. With the very soft
symmetry energy of $x=1$ (it first rises then starts decreasing with
the increasing $\rho$ above about $1.3\rho_0$~\cite{das03}, mimicking
one of the results in Ref.~\cite{wff}), however, the isospin
asymmetries at both very low and very high densities are higher than
the average asymmetry of the reaction system.  The observed inverse
relationship between the $\delta(\rho)$ and $E_{\mathrm{sym}}(\rho)$
is consistent with the well-known isospin fractionation phenomenon
first predicted based on the thermodynamics of asymmetric
matter~\cite{serot,liko}.  Similar to neutron skins in heavy nuclei,
neutron-rich clouds are dynamically generated in heavy-ion reactions
via the isospin diffusion. This indicates that the same underlying
physics is at work~\cite{steiner05b}.

In the following, we examine
the strength of the isospin diffusion and the thickness
of neutron skin in $^{208}$Pb as a function of the slope parameter
$L\equiv3\rho _{0}(\partial E_{\text{sym}}/\partial \rho)_{\rho_0}$.
\begin{figure}[tbh]
\begin{center}
\includegraphics[scale=0.80,angle=-90]{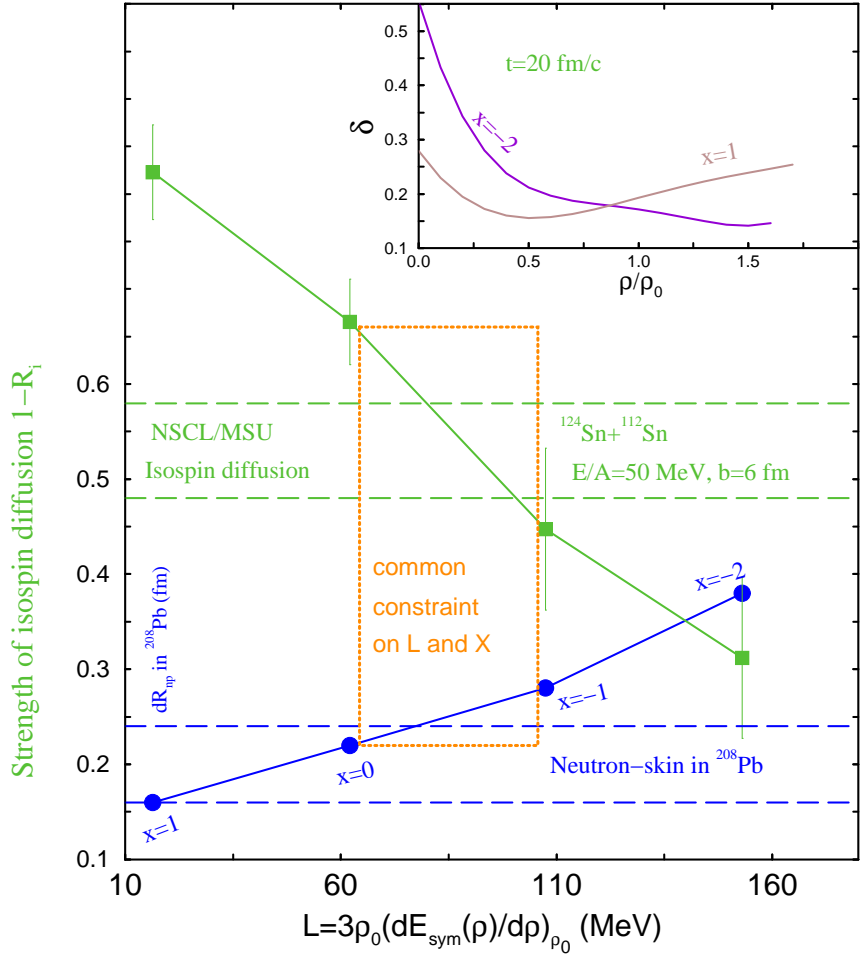}
\caption{The strength of isospin diffusion in the
$^{124}$Sn$+^{112}$Sn reaction and the size of neutron skin in
$^{208}$Pb as a function of the slope of the symmetry energy,
respectively.  The insert is the correlation of the isospin asymmetry
and density at the instant of 20 fm/c in the reaction considered.}
\end{center}
\label{figure2}
\end{figure}
The degree of isospin diffusion in the reaction of A+B is
experimentally measured by using~\cite{rami}
\begin{equation}
R_{i}\equiv\frac{2O_I^{A+B}-O_I^{A+A}-O_I^{B+B}}{O_I^{A+A}-O_I^{B+B}},
\end{equation}
where $O_I$ is any isospin-sensitive observable. The frequently used
ones include the neutron/proton ratio of pre-equilibrium nucleons,
ratios of light mirror nuclei and the isospin asymmetry of
projectile-like fragments. They all give essentially the same
result~\cite{betty04,bill}. By construction, the value of $R_{i}$ is
$1~(-1)$ for the symmetric $A+A~(B+B)$ reaction. If a complete isospin
equilibrium is reached in the asymmetric reaction $A+B$ as a result of
isospin diffusion the value of $R_{i}$ is about zero. The $R_i$ also
has the advantage of reducing significantly its sensitivity to the
symmetric part of the EOS. Shown in Fig. 2 are the strength of isospin
diffusion $1-R_i$ calculated using the transport model\cite{lichen}
and the size of neutron skin $dR_{np}$ in $^{208}$Pb calculated using
the Skyrme Hartree-Fock with interaction parameters adjusted such that
the same EOS is obtained~\cite{steiner05b}. The strength of isospin
diffusion $1-R_i$ is seen to decrease, while $dR_{np}$ increases with
the increasing $L$ as one expects. The NSCL/MSU data
$1-R_i=0.525\pm0.05$ implies that the $L (X)$ parameter is constrained
between 62.1 MeV (x=0) and 107.4 MeV (x=-1). This is consistent with
the meaurement $dR_{np}=0.2\pm 0.04$ fm~\cite{nskin} and also with
several recent
calculations~\cite{steiner05a,chuck,chuck2,jorge}. However, presently
available measurements of the neutron skin thickness using hadronic
probes have large systematic uncertainties associated with the strong
interaction.

\begin{figure}[tbh]
\begin{center}
\includegraphics[scale=0.40,angle=0]{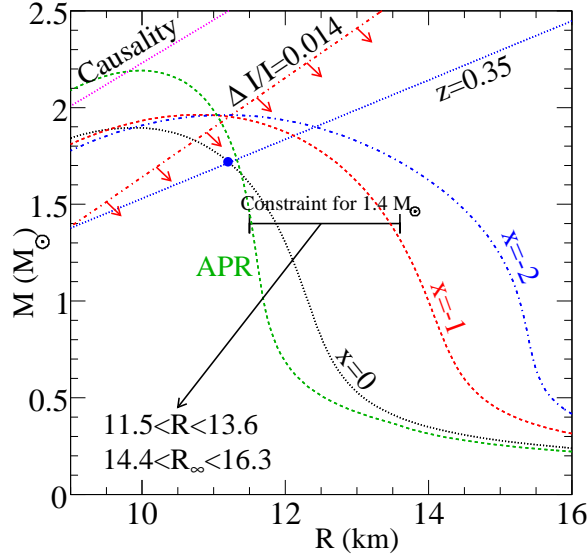}
\caption{{\protect\small The mass-radius curves for $x=0,-1,$ and $-2$
and the APR EOS. The limit from causality, the Vela pulsar, and the
redshift of EXO0748 are all indicated. The inferred radius of a 1.4
solar mass neutron star and the inferred value of $R_{\infty}$ are
given.}}
\end{center}
\label{figure3}
\end{figure}

The corresponding mass vs. radius curves for these EOSs, as well as
for APR (using the AV18+$\delta v$+UIX$^{*}$ interaction) are given in
Fig.~3.  In addition the constraints of causality, the mass-radius
relation from estimates of the crustal fraction of the moment of
inertia ($\Delta I/I=0.014$) in the Vela pulsar~\cite{link}, and the
mass-radius relation from the redshift measurement from
Ref.~\cite{cottam} are given. Any equation of state should be to the
right of the causality line and the $\Delta I/I$ line and should cross
the $z=0.35$ line. The horizontal bar indicates the inferred limits on
the radius and the radiation radius (the value of the radius which
is observed by an observer at infinity) defined as
$R_{\infty}=R/\sqrt{1-2GM/Rc^2}$ for a 1.4 \Msun~neutron star.

Since all three calculations with $x=0, -1$ and $x=-2$ have the same
compressibility ($K_0=211$ MeV) but rather different radii, it is
clear that the radius is indeed rather sensitive to the symmetry
energy while the maximum mass is only slightly
modified~\cite{lat01,lat01b,lat01c,prak01,pra88}.  The APR EOS has a
compressibility of $K_0=269$ MeV but almost the same symmetry energy
as with $x=0$. We note that the APR EOS leads to a 16\% higher maximum
mass ($1.9M_{\odot}$ to $2.2M_{\odot}$) but only a 5\% decrease in
radius (12.0 km to 11.5 km) as compared to the results with $x=0$.

Since only EOSs with symmetry energies between $x=0$ and $x=-1$ are
consistent with the isospin diffusion data and measurements of the
skin thickness of lead, we take them as representative of the possible
variation in neutron star structure that is consistent with
terrestrial data. The APR and the $x=0$ EOS have nearly identical
symmetry energies and slightly different radii. Neutron star radii are
strong functions of the symmetry energy but also contain contributions
from the isospin-symmetric part of the EOS, especially at higher
densities. Even though the compressibility of the APR EOS is larger
than that of the $x=0$ EOS, the pressure is typically lower in the APR
EOS at densities just above saturation, giving the APR EOS a smaller
radius by about 5\%. Thus we
take this 5\% difference as representative of the remaining
uncertainty in the symmetric part of the EOS and extend the minimum
radius to 11.5 km. Neutron stars with radii larger than 13.6 km are
difficult to make without a larger symmetry energy or
compressibility~\cite{steiner05a}.  We conclude that only radii
between 11.5 and 13.6 km (or radiation radii between 14.4 and 16.3 km)
are consistent with the $x=0$ and $x=-1$ EOSs, and thus consistent
with the laboratory data. It is interesting to note that a radius of
R=12.66 km was recently predicted for canonical neutron stars using a
new effective interaction calibrated by reproducing several collective
modes of $^{90}$Zr and $^{208}$Pb~\cite{jorge}. This radius falls
right in the range of our constraints. Our constraints on the radius
are also consistent with the range of radii from the extensive
analysis in Ref.~\cite{steiner05a} with only a few exceptions. The
field-theoretical models from this reference which are outside our
suggested range either have a relatively large symmetry energy at
saturation density ($\geq 36$ MeV), or have very soft symmetry
energies created by extremely strong non-linear couplings which are
atypical of most relativistic mean field models.

Our results suggest that the direct URCA processes is likely for stars
with masses larger than 1.39 \Msun, which is the limit obtained from
the $x=0$ EOS in Fig.~1. This constraint nearly matches the constraint
for the direct URCA process of 1.30 \Msun~obtained in
Ref.~\cite{jorge}. This is markedly different, however, from the
result from APR, which gives a large threshold for the direct URCA
process (even though the symmetry energy is very similar to our $x=0$
EOS).

{\it Does this constraint agree with present neutron star radius
observations?} The answer to this question is ``yes''. Assuming a mass
of 1.4 \Msun, the inferred radiation radius, $R_{\infty}$, (in km) is
$13.5\pm 2.1$~\cite{bob02,bob02b} or $13.6\pm 0.3$~\cite{gendre02a}
for the neutron star in $\omega$ Cen, $12.8\pm 0.4$ in
M13~\cite{gendre02b}, $14.5^{+1.6}_{-1.4}$ for X7 in 47
Tuc~\cite{ryb05} and $14.5^{+6.9}_{-3.8}$ in M28~\cite{becker},
respectively. Except the neutron star in M13 that has a slightly
smaller radius, all others fall into our constraints of 14.4 km
$<R_{\infty}< 16.3$ km within the observational error bars that are
often larger than the range we gave.
           
While the Vela $\Delta I/I$ upper limit does not provide any new
information, the fact that the $z=0.35$ line does not cross our range
of radii implies a mass larger than 1.4 \Msun~for EXO-0748~(the
minimum mass would be about 1.7 \Msun~corresponding to the dot in
Fig. 3). This is larger than the canonical 1.4 \Msun~ neutron star
mass, but is not unreasonable since this object is
accreting~\cite{cottam}\footnote{In fact, while revising this work,
a measurement of the mass 2.10 $\pm$ 0.28 $\mathrm{M_{\odot}}$ 
and radius 13.8 $\pm$ 1.8 km for this object was reported
in Ref.~\cite{ozel}, confirming that this object is likely more
massive than 1.4 $\mathrm{M_\odot}$.}.

While estimates of radii based on astrophysical observations are still
very challenging, it is useful to compare our results with recent
Chandra/XMM-Newton observations. Together with more refined
observations, future heavy-ion experiments (some recent progress in
Ref.~\cite{mike}) with advanced radioactive beam facilities~\cite{ria}
and measurements of parity violating electron-nucleus
scattering~\cite{parity} will allow us to pin down more precisely the
EOS of neutron rich matter. This would allow tighter contraints on
neutron star radii. On the other hand, a neutron star radius
measurement outside of our prediction may indicate non-standard
physics.

We would like to thank Lie-Wen Chen and Sanjay Reddy for helpful
discussions and the anonymous referees for their comments. The work of
B.A. Li was supported in part by the NSF under Grant No. PHY-0354572,
PHY0456890 and the NASA-Arkansas Space Grants Consortium Award
ASU15154. The work of A.W. Steiner was supported by the DOE under
grant no. DOE/W-7405-ENG-36.

\end{document}